\documentclass{article}
\usepackage[utf8]{inputenc}
\usepackage[margin=0.9in]{geometry}
\usepackage{amsmath}
\usepackage{amssymb}
\usepackage{float}
\usepackage{mathtools}
\usepackage{subfig}
\usepackage{authblk}
\usepackage{verbatim}
\usepackage{physics}
\usepackage{enumitem}
\usepackage{float}
\usepackage{qcircuit}

\usepackage[backend=biber, sorting=none]{biblatex}
\usepackage[colorlinks,citecolor=blue]{hyperref}
\usepackage{hyperref}
\usepackage{dcolumn}
\usepackage[table]{xcolor}
\hypersetup{
    colorlinks=true,
    linkcolor=red,
    filecolor=blue,      
    urlcolor=blue,
}

\title{\textbf{Simulating quantum chaos on a quantum computer}}
\author[1,3,4]{Amit Anand}

\affil[1]{Department of Mechanical Engineering\\ Indian Institute of Engineering Science And Technology\\ Shibpur, Howrah-711103,West Bengal, India.}

\author[2,3,4]{Sanchit Srivastava}

\affil[2]{School of Physics, Indian Institute of Science Education and Research, Thiruvananthapuram - 695551, Kerala, India}
\author[3,4]{Sayan Gangopadhyay}
\affil[3]{Department of Physics and Astronomy, University of Waterloo, Waterloo, Ontario, Canada N2L 3G1 }
\author[4,5]{Shohini Ghose}

\affil[4]{Institute for Quantum Computing, University of Waterloo, Waterloo, Ontario, Canada N2L 3G1}
\affil[5]{Department of Physics and Computer Science, Wilfrid Laurier University, Waterloo, Ontario, Canada N2L 3C5}

\date{}
\begin{document}

\maketitle

\begin{abstract}
  We show that currently available noisy intermediate-scale quantum (NISQ) computers can be used for versatile quantum simulations of chaotic systems. 
  We introduce a novel classical-quantum hybrid approach for exploring the dynamics of the chaotic quantum kicked top (QKT) on a  universal quantum computer. The programmability of this approach allows us to experimentally explore the complete range of QKT chaoticity parameter regimes inaccessible to previous studies. Furthermore, the number of gates in our simulation does not increase with the number of kicks, thus making it possible to study the QKT evolution for arbitrary number of kicks without fidelity loss. Using a publicly accessible NISQ computer (IBMQ), we  observe periodicities in the evolution of the 2-qubit QKT, as well as signatures of chaos in the time-averaged 2-qubit entanglement. We also demonstrate a  connection between entanglement and delocalization in the 2-qubit QKT, confirming theoretical predictions.

 \end{abstract}


\section{Introduction}

The seeds that grew into the field of quantum computing were initially sown in the context of quantum simulations\cite{Cirac_Zoller_2012,Altman_Brow}. As Feynman pointed out\cite{Feynman_1986}, the possibility of mapping one quantum system into another opens up new windows for efficiently exploring the properties and dynamics of general quantum systems\cite{Leontica_Tennie_Farrow_2021,García-Pérez_Rossi_Maniscalco_2020}. Large-scale, programmable quantum computers could offer exactly this type of possibility of mapping and simulating complex quantum systems\cite{RevModPhys.86.153}. While such large-scale quantum computers do not yet exist, it is worth exploring the potential for quantum simulations using currently available noisy intermediate-scale quantum (NISQ) computers\cite{Preskill_2018,McArdle_Endo_Aspuru-Guzik_Benjamin_Yuan_2020}. One such potential area of NISQ application is the topic of quantum chaos - the study of quantum systems that exhibit chaos in some classical limit. The question of how classical chaos emerges from quantum dynamics remains one of the open fundamental questions in quantum theory\cite{Zurek_Paz_1995,haake_1987_classical,haake_1987_classical,haake1991book}. On the experimental side as well, the quantum control and precision needed to explore quantum chaotic dynamics over a wide range of parameters and long time scales remains quite challenging. So far, relatively few experiments in limited parameter regimes and for short times have been performed\cite{Li_Fan_Wang_Ye_Zeng_Zhai_Peng_Du_2017,Szriftgiser_Lignier_Ringot_Garreau_Delande_2003,NMR,Shohini2009,Google}. In this work, we demonstrate the use of NISQ computers for flexible simulations of quantum chaos.

Classical chaos is characterized by exponential sensitivity to initial conditions quantified by the Lyapunov exponent that is a measure of the rate of divergence of neighbouring trajectories \cite{Smith1995TheGP,Ott_2002,DATTA_2004}. A corresponding quantum measure of chaos is challenging to define due to the uncertainty principle and the linearity of quantum evolution. In recent years, the question of quantum-classical correspondence in classically chaotic systems has been explored in the context of quantum information processing. The connection between fundamentally quantum phenomena such as entanglement and classical chaos has puzzled physicists for decades, and has gained new relevance for quantum computing applications. 
It has been shown that classical chaos can affect the implementation of quantum computing algorithms \cite{PhysRevE.62.3504,PhysRevE.62.6366}. Chaos can also affect the generation of dynamical entanglement, an important resource for quantum computing. 

To understand chaos in the quantum context, it is important to explore signatures of classical chaos in the deep quantum regime, where the standard Bohr  correspondence principle cannot be invoked. A textbook model for studying quantum chaos is the quantum kicked top \cite{haake_1987_classical}, which is a finite-dimensional spin system that displays chaotic dynamics in the classical limit. The quantum kicked top (QKT) has been extensively studied theoretically \cite{discord,dogra_google,EntSig,ghose_paul_stock,ghose_sanders,ghose2008chaos,pattanayak,Santhanam}. The system can be described as a collection of indistinguishable qubits, which makes it attractive to explore in the framework of quantum information processing and NISQ devices. In the deep quantum regime, periodicities and symmetries in the two- and three-qubit QKT model were theoretically studied \cite{Santhanam,pattanayak}. A few experimental studies of the QKT have also been performed \cite{NMR,Shohini2009,Google}. In \cite{Google}, a 3-qubit model of the QKT was shown to exhibit ergodic dynamics and a resemblance between entanglement entropy and classical phase space dynamics was noted. Temporal periodicity and symmetries of the 2-qubit QKT were explored using NMR techniques in \cite{NMR}. 
These experiments are limited to a small number of kicks due to decoherence times of the physical qubits. Furthermore, in these implementations the chaoticity parameter $\kappa$ is determined by the strength and duration of interaction between the qubits, making it difficult to tune.
To experimentally study the long term dynamics and dependence on $\kappa$ rigorously, one needs to explore longer time scales and a wider range of $\kappa$. In this work, we show that mapping the QKT on to a programmable quantum circuit in a quantum computer allows simulations of the QKT that overcome previous experimental limitations. This opens new regimes of experimental exploration in both time and in parameter space.

We construct and demonstrate for the first time, an exact simulation of the 2-qubit quantum kicked top using a universal set of quantum logic gates. Our quantum circuit-based simulation is programmable and enables flexible initial state preparation and evolution. Using IBM's 5-qubit chip $vigo$, we can prepare initial states and implement the dynamics of the QKT for an arbitrary number of kicks and a wide range of $\kappa$. The number of gates required for this simulation is independent of the number of kicks and  value of $\kappa$. Therefore, our model does not suffer any systematic loss in fidelity with increasing number of kicks or $\kappa$ values. Finally, full quantum state tomography enables us to explore signatures of chaos in 2-qubit entanglement.

The ability to vary $\kappa$ and the number of kicks allows us to experimentally observe the periodic nature of the dynamics with respect to $\kappa$ as well as kick number. Additionally, the temporal periodicity of the QKT can be used to obtain highly accurate time averages of relevant physical quantities. In particular, we explore the time-averaged  entanglement for different initial spin coherent states (SCS).  We find that a contour plot of the time average entanglement shows clear signatures of the classical phase space structures of regular islands in a chaotic sea, even in a deep quantum regime. We also show that the states initialized in chaotic regions of the phase space show intermediate values of average concurrence, whereas, the fixed points and the period-4 orbit correspond to the minimum and maximum values respectively. This behaviour is related to the degree of delocalization of the state and thus demonstrates a connection between delocalization and entanglement \cite{kumari_2019_quantumclassical}.

Our work shows that current quantum computers are useful for flexibly exploring new experimental regimes in quantum chaotic systems. Mapping the system onto a tunable quantum circuit lets us probe different aspects of the QKT dynamics without the need for building sophisticated customized hardware or being constrained by fixed system parameters. This method combines the ease of numerical simulation with the built-in quantum evolution of a physical system.

The paper is organized as follows: In section \ref{sec 2}, the the quantum kicked top model is introduced and the chaotic classical limit is discussed. In section \ref{sec 3}, our circuit-based approach and the mapping to the IBM $vigo$ processor is described. In section \ref{sec 4}, the experimental results of the IBM $vigo$ simulations are discussed. Finally, conclusions, outlook and the scope of application of our circuit-based approach are discussed in section \ref{sec 5}.


\section{The quantum kicked top}\label{sec 2}

The quantum kicked top (QKT) model was first introduced by Haake, Kus and, Scharf in 1987 \cite{haake1987}. It is now a widely studied model for exploring quantum chaos. The QKT is a time-dependent periodic spin system governed by the  Hamiltonian 

\begin{equation}
H = \hbar\frac{p J_{y}}{\tau} +  \hbar \frac{\kappa J_{z}^2}{2j}\Bigg( \sum_{n=-\infty}^{n=\infty} \delta (t-n\tau) \Bigg),
 \end{equation}
where \emph{ $J_{x},J_{y}$} and \emph{$J_{z}$} obey the standard angular momentum operator commutation relations. The operator \emph{$J^{2}=J(J+1)\hbar^{2}$} commutes with the Hamiltonian and the magnitude of the angular momentum is a constant of motion. The Hamiltonian consists of a series of rotations (kicks) described by the \emph{$J_{y}$} term alternating with torsion due to the non-linear \emph{$J^{2}_{z}$} term. $\tau$ is the duration between kicks, \emph{p} is the angle of linear rotation in the y-direction, and $\kappa$ is the strength of twist in the z-direction. The kick-to-kick Floquet time evolution operator can be written as
\begin{equation}
    U= \exp\Bigg(-i\frac{\kappa }{2j}J_{z}^2\Bigg)\exp\Bigg(-i\frac{p }{\tau}J_{y}\Bigg).
\end{equation}

The classical map for the kicked top can be obtained by writing the Heisenberg equations of motion for the angular momentum operators and then taking the limit ${ \emph{j} \to \infty}$. By defining the normalized variables \emph{$X=J_{x}/j$},   \emph{$Y=J_{y}/j$} and  \emph{$Z=J_{z}/j$}, the classical equations of motions for \emph{p}=$\pi/2$ are
\begin{eqnarray}
 \nonumber X_{n+1}(\tau)&=&Z_n(\tau)\cos(\kappa X_n(\tau))+Y_n(\tau)\sin(\kappa X_n(\tau)), \\ 
 \nonumber Y_{n+1}(\tau)&=&Y_n(\tau)\cos(\kappa X_n(\tau))-Z_n(\tau)\sin(\kappa X_n(\tau)), \\
 Z_{n+1}(\tau)&=&-X_n(\tau)
\end{eqnarray}
Here the dynamical variables (X,Y,Z) satisfy the constraint $X^{2} + Y^{2} + Z^{2} =1$, i.e., they are restricted to be on the unit sphere $\emph{S}^2$. Thus, the variables can be parameterized into spherical polar coordinates as $(X,Y,Z)=(\sin(\theta)\cos(\phi), \sin(\theta)\sin(\phi),\cos(\theta))$. 

\begin{figure}[!h]
  \centering
  \subfloat[]{\includegraphics[width=0.5\textwidth]{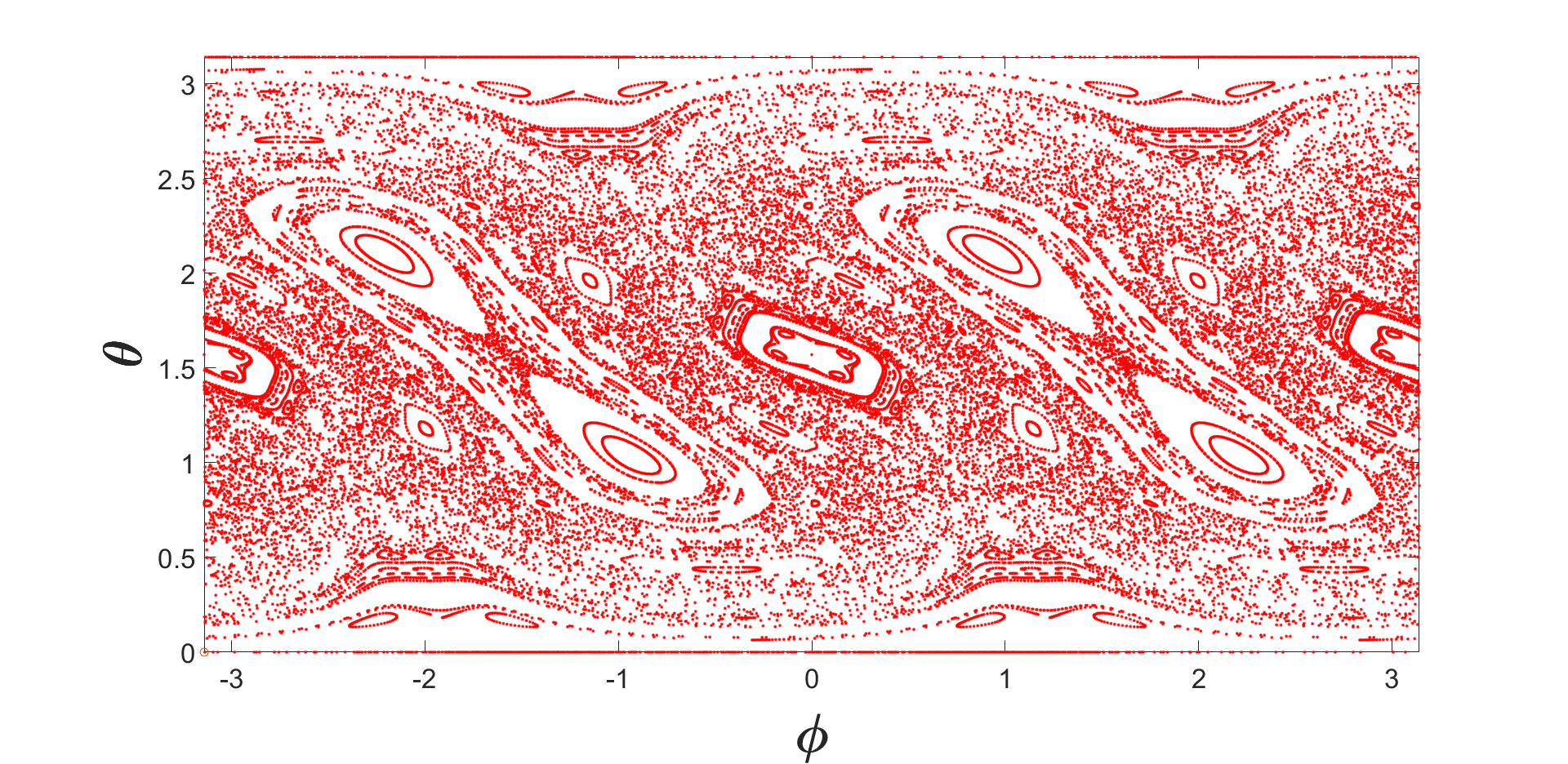}\label{classical_k2.5}}
  \hfill
  \subfloat[]{\includegraphics[width=0.5\textwidth]{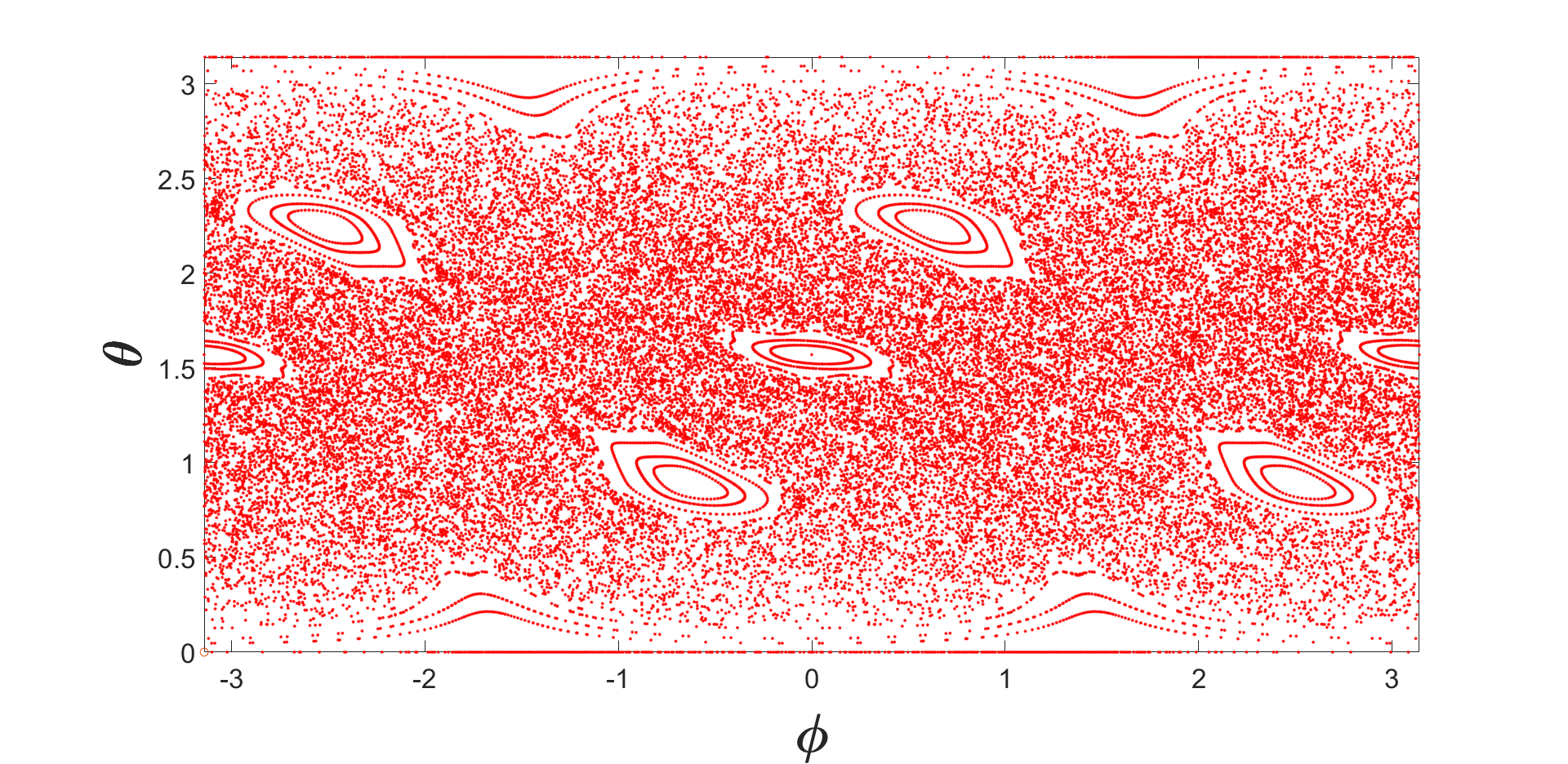}\label{classical_k3}}
  \caption{ Stroboscopic map showing the classical time evolution over 150 kicks for \textbf{a}. k=2.5 and \textbf{b}. k=3.0 for 289 initial points in phase space.} 
\end{figure}

As the  chaoticity parameter $\kappa$ is varied , the classical dynamics ranges from regular motion (for $\kappa$ $\leq $ 2.1), to a mixture of regular and chaotic behaviour for different  initial conditions (for 2.1 $\leq \kappa \leq 4.4$), to fully chaotic motion (for $\kappa > 4.4$). The classical stroboscopic map (in spherical co-ordinates) for a range of initial conditions with $\kappa$ = 2.5 and $\kappa$ =3 is plotted in Fig. \ref{classical_k2.5} and Fig. \ref{classical_k3} respectively.

\section{Simulating quantum kicked top dynamics} \label{sec 3}

\subsection{Floquet operator}

The time evolution of the system is described by evolving the initial state with the Floquet time evolution operator 
\begin{eqnarray}
  U = \exp\Bigg(-i\frac{\kappa }{2j}J_{z}^2\Bigg) \exp\big(-ipJ_{y}\big).
\end{eqnarray}
Since $[H,J^2] = 0$ for the QKT Hamiltonian, it can be considered as an $N = 2j$ qubit system \cite{kumari_2019_quantumclassical}. The spin-$j $ operators are written in terms of the single qubit Pauli rotation operators as:  
\begin{equation}
J_{\alpha}=\frac{1}{2} \sum_{i=1}^{2 j} \sigma_{i \alpha}, \quad \alpha \in\{x, y, z\}
\end{equation}
where $\sigma_{i\alpha}$ denotes $\sigma_{\alpha}$ acting on the $i^{th}$ qubit. This allows us to rewrite the QKT Hamiltonian in 2j-qubit space: 
\begin{equation}
H=\hbar \frac{\kappa}{8 j}\left(2 j+\sum_{i, k=1 \atop i \neq k}^{2 j} \sigma_{i z} \otimes \sigma_{k z}\right) \sum_{n=-\infty}^{\infty} \delta(t-n \tau)+\hbar \frac{p}{2 \tau} \sum_{i=1}^{2 j} \sigma_{i y}.
\end{equation}
In particular, the $j=1$ QKT is described by the 2-qubit Hamiltonian 

\begin{equation}
    H =  \frac{\hbar k}{4}\left( \mathbb{I} + \sigma_z \otimes \sigma_z\right)\sum_{n}\delta(t-n\tau) + \frac{\hbar p}{2\tau}(\sigma_y\otimes\mathbb{I}+\mathbb{I}\otimes\sigma_y)
\end{equation}
and, setting $\hbar=1$, the corresponding single-kick time evolution unitary is

\begin{equation}
    U = \exp\bigg(-i\frac{\kappa}{4}(\mathbb{I}+\sigma_z \otimes \sigma_z)\bigg)\exp\bigg(-i\frac{p}{2}(\sigma_y\otimes\mathbb{I}+\mathbb{I}\otimes\sigma_y)\bigg).
\end{equation}

\subsection{Initial States}
Minimum uncertainty states in spin systems are spin coherent states (SCS)
\cite{radcliffe_1971_some} that satisfy the uncertainty relation 

\begin{equation}
    \Delta J_i\Delta J_k = \frac{\hbar}{2}|\Delta J_l|,
\end{equation}
where $i$, $k$ and $l$ are permutations of x,y and z. The uncertainty for these states is distributed symmetrically over the two operators. For larger $j$ values, the SCS becomes highly localized around the point $(\theta,\phi)$ in the phase space and hence in the classical limit of $j\rightarrow\infty$, approximates the classical angular momentum state located at $(\theta,\phi)$ \cite{haake_1987_classical}. 

Spin squeezed states, which have asymmetric distribution of uncertainty, can display entanglement in the corresponding multi-qubit representation\cite{ma_2011_quantum}. Since we are interested in studying entanglement that arises from the dynamics of the system, we choose SCSs as our initial states, thus ensuring there is no initial entanglement between the qubits.  
Given any point $(\theta,\phi)$ in the classical phase space, we construct the corresponding SCS $\ket{j;\theta,\phi}$ 
\begin{equation}
    \ket{j;\theta,\phi} = \exp[i\theta(J_x\sin\phi-J_y\cos\phi)]\ket{j,j}. 
\end{equation}
In the $2j$-qubit space, we define our initial states as the SCSs
\begin{equation}
     \ket{j;\theta,\phi} = \ket{\theta,\phi}^{\otimes2j},
\end{equation}
where $\ket{\theta,\phi}$ are points on the Bloch sphere.

\subsection{Implementation of unitaries with a quantum circuit}

Any n-qubit unitary can be decomposed into $2^n(2^n-1)/2$ single-qubit gates with controls. We follow the prescription in \cite{li_2013_decomposition} to decompose our 2-qubit Floquet operator as 
\begin{equation}
    U=U_{1}\times U_{2}\times U_{3}\times U_{4}\times U_{5}\times U_{6},
\end{equation} 
where $U_i$ are either single-qubit unitaries or controlled unitaries of the form 
\begin{eqnarray}
\nonumber U_i\equiv\Qcircuit @C=1em @R=.7em {
& \gate{V_i} & \qw \\
& \ctrl{-1} & \qw  
}
& \text{or} & U_i \equiv\Qcircuit @C=1em @R=.7em {& \ctrl{1} & \qw \\
& \gate{V_i} & \qw 
}.
\end{eqnarray}
In the circuit above, the two wires represent the two qubits, the solid dot represents the control qubit, and the box indicates the single qubit operation $V_i$ on the target qubit. 
The exact decomposition for a 2-qubit gate in this scheme is 
\begin{equation} \label{U gates}
    U = \Qcircuit @C=1em @R=.7em {
    & \ctrl{1} & \gate{V_5} & \ctrl{1} & \gate{V_3} & \ctrl{1} & \qw & \qw \\
    & \gate{V_6} & \ctrl{-1} & \gate{V_4} & \qw & \gate{V_2} & \gate{V_1} & \qw &.  \\
    } 
\end{equation}

Implementation of these gates on a quantum computer requires further decomposition into rotations and CNOT gates. Given a 1-qubit unitary W$\in$ SU(2), a controlled gate of the from ($\ket{0}\bra{0}\otimes\mathbb{I}+\ket{1}\bra{1}\otimes W)$ can be decomposed as 
\begin{equation}\label{cw gates}
\Qcircuit @C=1em @R= 1em {& \ctrl{1} & \qw \\
& \gate{W} & \qw 
} = \Qcircuit @C=0.7em @R=.7em {
& \qw  & \ctrl{1} & \qw & \qw & \ctrl{1}&  \qw & \qw &\qw \\
& \gate{R_z(\frac{\beta-\alpha}{2})} & \targ & \gate{R_z(-\frac{\alpha+\beta}{2})} & \gate{R_y(-\frac{\theta}{2})} & \targ & \gate{R_y(\frac{\theta}{2})} & \gate{R_z(\alpha)} & \qw }
\end{equation}
where $R_x$, $R_y$ and $R_z$ describe rotations on the Bloch sphere and $\alpha$, $\beta$ and $\theta$ are such that\\ $R_z(\alpha)R_y(\theta)R_z(\beta) = W$.

A general 1-qubit unitary V is of the form $\exp(i\delta)\times W$ where W$\in$ SU(2). A 2-qubit gate of the form ($\ket{0}\bra{0}\otimes\mathbb{I}+\ket{1}\bra{1}\otimes V)$ can be written as 
\begin{equation} \label{cv gates}
\Qcircuit @C=1em @R=.7em {
& \ctrl{1} & \qw \\
& \gate{V} & \qw 
} = \Qcircuit @C=1em @R=.7em {
& \ctrl{1} & \ctrl{1} & \qw  \\
& \gate{W} & \gate{U_{\delta}} & \qw 
}
\end{equation}
where $U_{\delta} = \exp(-i\delta)\times\mathbb{I}$. This controlled phase gate can be further simplified by moving the phase over to the other qubit: 
 \begin{eqnarray}
 \nonumber(\ket{0}\bra{0}\otimes \mathbb{I} + \ket{1}\bra{1}\otimes U_{\delta})&=& (\ket{0}\bra{0}\otimes \mathbb{I}+ \exp(i\delta)\ket{1}\bra{1}\otimes \mathbb{I}) \\ \nonumber
 &=& (\ket{0}\bra{0} + \exp(i\delta)\ket{1}\bra{1})\otimes\mathbb{I}\\
 &=& R_z(\delta)\otimes\mathbb{I}. 
 \end{eqnarray}
Here, we ignore a global phase factor of $\exp(i\delta/2)$ in the final step. 
Hence, we are left with 
\begin{equation}
\Qcircuit @C=1em @R= 1.6 em {& \ctrl{1} & \qw \\
& \gate{V} & \qw 
} = \Qcircuit @C=0.7em @R=.7em {
& \qw  & \ctrl{1} & \qw & \qw & \ctrl{1}& \gate{R_z(\delta)} & \qw &\qw \\
& \gate{R_z(\frac{\beta-\alpha}{2})} & \targ & \gate{R_z(-\frac{\alpha+\beta}{2})} & \gate{R_y(-\frac{\theta}{2})} & \targ & \gate{R_y(\frac{\theta}{2})} & \gate{R_z(\alpha)} & \qw  & .}
\end{equation}

Similar analysis follows when the control and target qubits are exchanged. For two-qubit unitaries of the type $\mathbb{I}\otimes V$ and $V\otimes\mathbb{I}$, the phase factors appearing on $V$ are global and can be ignored. These gates have a similar decomposition as the one in Eq.~\ref{cw gates} with the CNOT gates replaced by X gates. For example, a unitary of the form $\mathbb{I}\otimes V$ can be decomposed as 
\begin{equation}
\Qcircuit @C=1em @R= 1em {& \qw & \qw \\
& \gate{V} & \qw 
} = \Qcircuit @C=0.7em @R=.7em {
& \qw  & \qw & \qw & \qw & \qw &  \qw & \qw &\qw \\
& \gate{R_z(\frac{\beta-\alpha}{2})} & \gate{X} & \gate{R_z(-\frac{\alpha+\beta}{2})} & \gate{R_y(-\frac{\theta}{2})} & \gate{X} & \gate{R_y(\frac{\theta}{2})} & \gate{R_z(\alpha)} & \qw & .}
\end{equation}

In this scheme, our 2-qubit Floquet operator can be constructed from 46 total gates, with 8 two-qubit CNOT gates and 38 single qubit rotations. Consecutive rotations have been counted as separate single qubit gates here. Depending on the universal gate set for the particular quantum computer, the actual number of gates needed to simulate the Floquet operator unitary may be reduced. 

\subsection{Decomposition into $U_1$ and $U_3$ gates on IBMQ}

We implement our quantum circuits on the quantum hardware and simulator back end of the IBM Quantum Experience \cite{ibm}. The interfacing with the quantum hardware is done using Qiskit \cite{qiskit}. 

Qiskit allows us to implement 1-parameter and 3-parameter single-qubit unitary operators of the form

\begin{equation}
\nonumber U_3(\theta,\phi,\lambda)=\left(\begin{array}{cc}
\cos (\theta / 2) & -e^{i \lambda} \sin (\theta / 2) \\
e^{i \phi} \sin (\theta / 2) & e^{i \lambda+i \phi} \cos (\theta / 2) 
\end{array}\right)
\end{equation}

\begin{equation}
\nonumber U_1(\lambda)=\left(\begin{array}{cc}
1 & 0 \\
0 & e^{i \lambda}  
\end{array}\right). 
\end{equation}

We decompose the gates in Eq.\ref{U gates} into a combination of $U_1$ and $U_3$ gates. Gates of the form $V\otimes\mathbb{I}$ and $\mathbb{I}\otimes V$ are implemented directly as $U_3$ gates. For controlled gates, the decomposition given in Eq.\ref{cv gates} is used where W$\in$ SU(2) is implemented as a $U_3$ gate and $U_{\delta}$ is implemented as $U_1(\delta)$. Hence, we obtain the final circuit decomposition for our 2-qubit Floquet operator on IBMQ:

\begin{equation}\label{U final}
  U =  \Qcircuit @C=1em @R= 1em {
   & \ctrl{1} & \gate{U_1(\delta_6)} & \gate{W_5} & \qw & \ctrl{1} & \gate{U_1(\delta_4)} & \gate{W_3} & \ctrl{1} & \gate{U_1(\delta_2)} & \qw \\ 
   & \gate{W_6} & \qw & \ctrl{-1} & \gate{U_1(\delta_5)} & \gate{W_4} & \qw & \qw & \gate{W_2} & \gate{W_1} & \qw 
   } 
\end{equation}
with $W_i = U_3(\theta_i,\phi_i, \lambda_i)$. 

Time evolution after multiple kicks is calculated by applying the Floquet unitary on the initial state repeatedly. This could be achieved by applying the set of gates given in Eq.\ref{U final} consecutively $N$ times to simulate evolution by $N$ time steps. However, to mitigate the errors which may arise from the increasing number of gates, in our approach, we decompose the effective N-step unitary $U^N$ using the same procedure as mentioned above. This means that the state after any arbitrary number of steps can be obtained by applying the same set of gates given in Eq.~\ref{U final} with appropriate parameters. The advantage of this approach is that the number of gates is fixed and does not grow with the number of kicks. Similarly, time evolution for different values of $\kappa$ can be implemented by computing the relevant parameters for the set of gates corresponding to the unitary $U^N(\kappa)$. This affords us a more fine-grained and flexible control over this parameter compared to other qubit-based realizations where the value of $\kappa$ is set by tuning the time duration of interactions between the physical qubits. 

After applying the appropriate set of gates to the initial states, the final states density matrix is constructed using state-tomography circuits built into Qiskit Ignis\cite{Aleksandrowicz}. Physical quantities of interest can be calculated from this density matrix.  



\section{Exploring quantum chaos with a quantum computer} \label{sec 4}

\subsection{Performance of quantum simulations}

\begin{figure}[!h]
\centering
\includegraphics[width=1.1\textwidth]{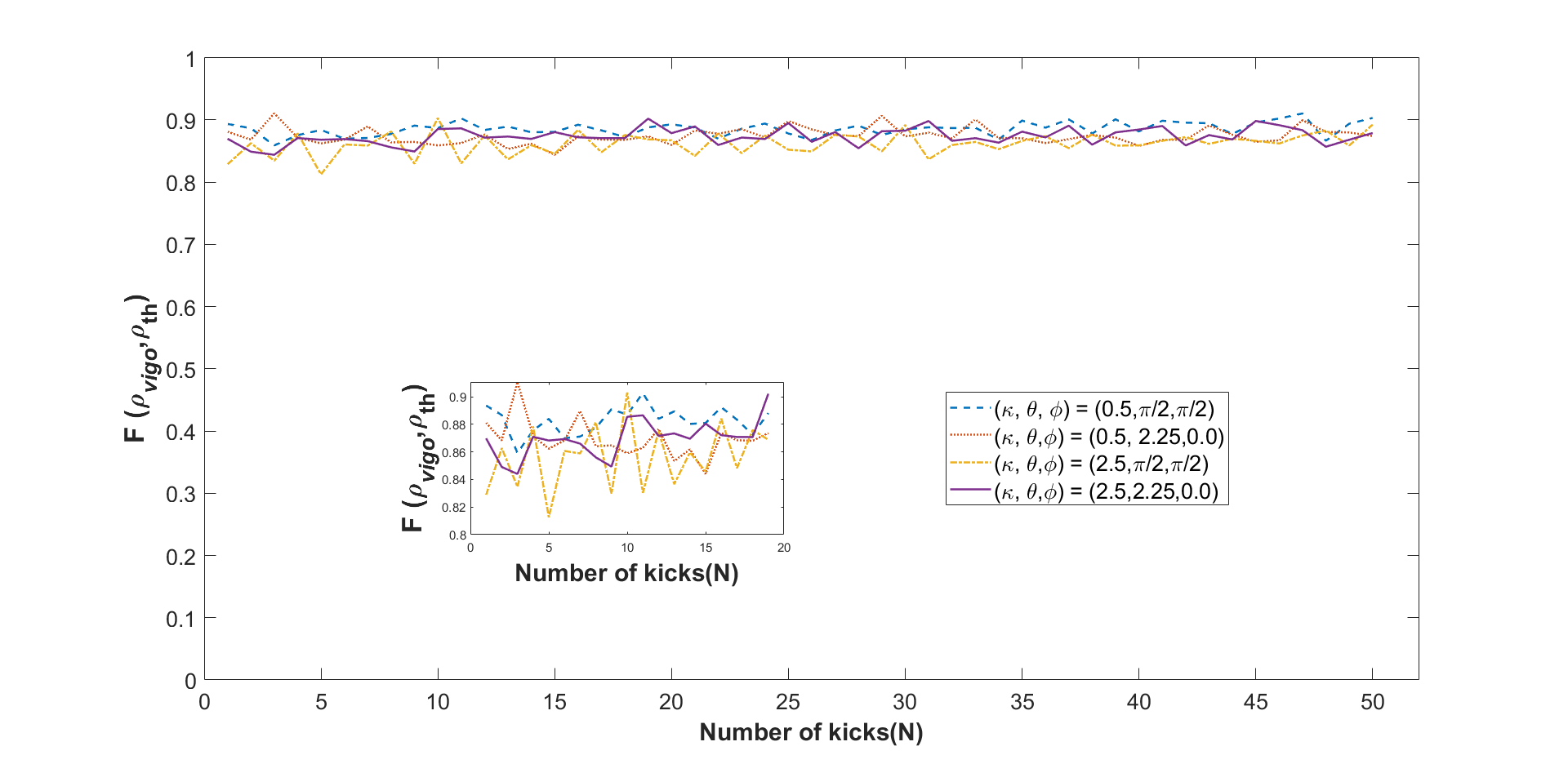}
  \caption{Fidelity of the tomographically reconstructed 2-qubit state for different initial states and different $\kappa$ values on IBM \textit{$vigo$}}
  \label{FIG1}
  \end{figure}
\begin{figure}[!h]
\centering
\includegraphics[width=1.1\textwidth]{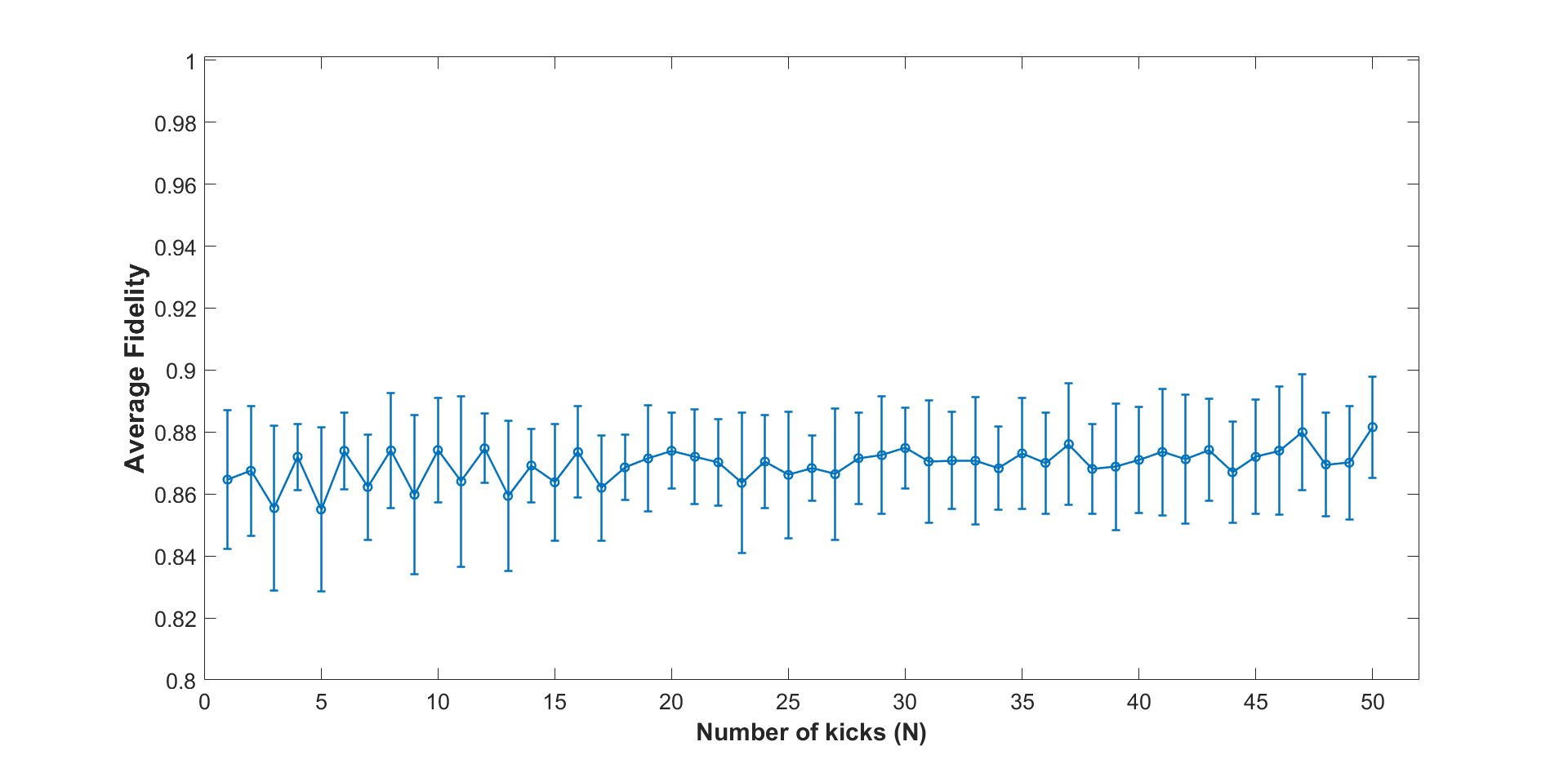}
  \caption{Fidelity of the tomographically reconstructed 2-qubit state averaged over initial states with 
   $(\theta,\phi)\in\{(2.25,0),(\pi/2,\pi/2),(\pi/2,0)\}$ and $\kappa\in\{0.5,2.5,4.5,6.5\}$ on IBM \textit{$vigo$}. The error bars indicate standard deviation. }
  \label{FIG2}
\end{figure}

Starting with various initial points for two different values of $\kappa$, we apply the quantum circuit for implementing $N$ kicks on IBM $vigo$. We reconstruct the resulting final state by performing quantum state tomography and use the fidelity of the reconstructed state as a measure of simulation accuracy. For the theoretically predicted state $\rho_{\text{th}}$ and the reconstructed state $\rho_{\text{$vigo$}}$, fidelity is given by $F(\rho_{\text{$vigo$}},\rho_{\text{th}})=(\text{tr}(\sqrt{\sqrt{\rho_{\text{th}}}\rho_{\text{$vigo$}}\sqrt{\rho_{\text{th}}}} ))^2$. We observed that there is no systematic loss in fidelity with the number of kicks for different initial states and values of $\kappa$ (Fig. \ref{FIG1}). We can compute the average fidelity over different $\kappa$ values in the range [0.5,6.5], and for different initial states as a function of time. As seen in Fig. \ref{FIG2}, the average fidelities remain around 0.87.

To obtain a benchmark of how well our circuit method implemented on a publicly accessible quantum computer (IBM $vigo$) performs, we have compared the observed fidelities with that of previous QKT experiments\cite{Google}, \cite{NMR}. \cite{Google} reported monotonically decreasing fidelity with a low of 0.6 for 10 kicks. \cite{NMR} reported a significant drop in fidelity from $6^{th}$ to $8^{th}$ kick with the lowest fidelity being 0.8. In our study, the non-decreasing trend in fidelity for much higher number of kicks can be attributed to the fixed number of gates for arbitrary number of kicks. By decomposing the unitary into elementary, programmable quantum gates, we effectively remove any constraints on the parameters of the physical system (QKT) that we can implement. In the IBM-Q systems, the error varies only with the number of single qubit physical rotations and CNOT gates \cite{nishio_2020_extracting} acting on each qubit. Since the number of gates in the circuit remains constant irrespective of the value of $\kappa$, we observe that the fidelity values do not depend on $\kappa$.

\subsection{Observation of periodicity in $\kappa$}

\begin{figure}[!h]
\centering
\includegraphics[width=1.0\textwidth]{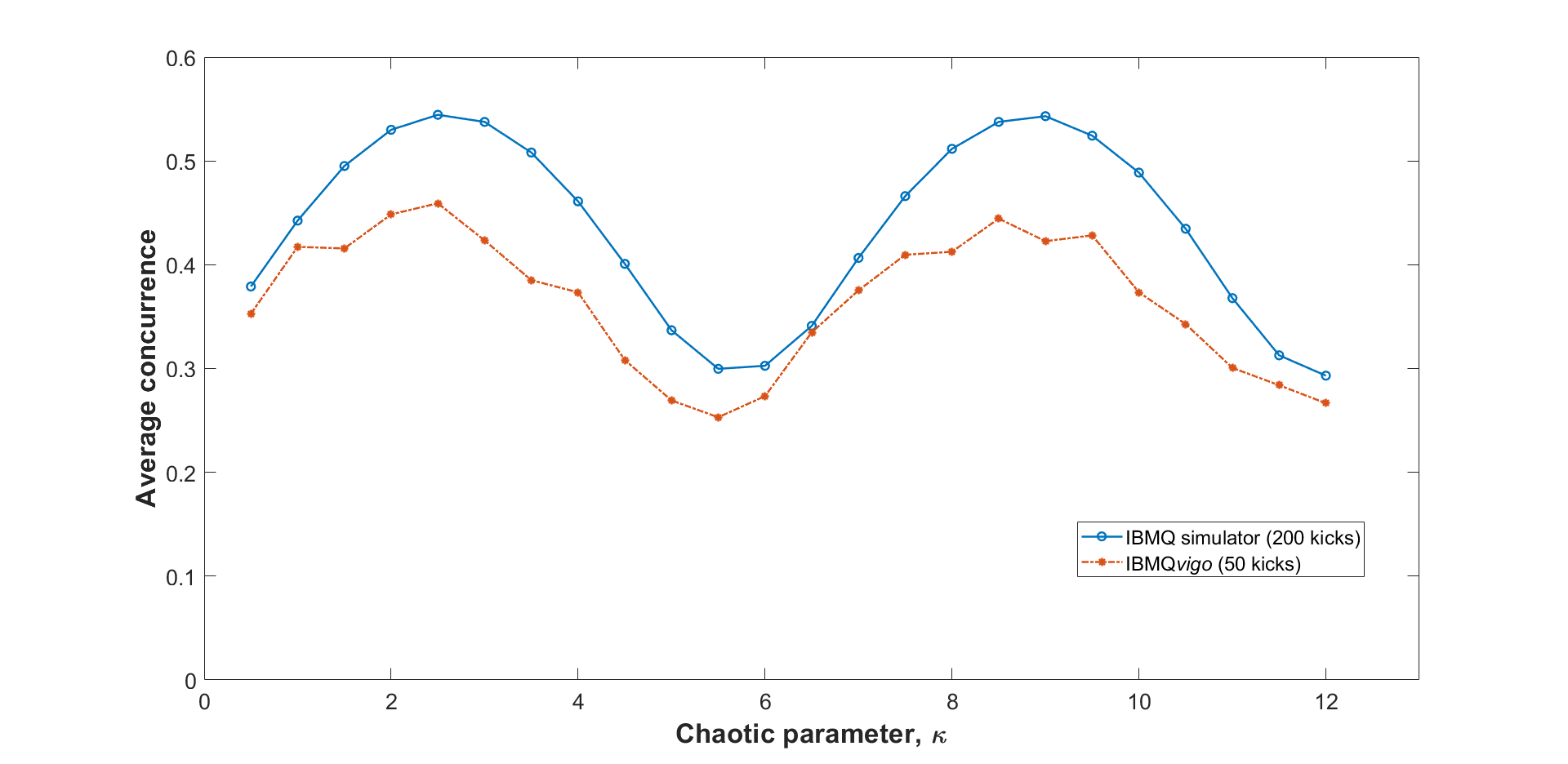}
  \caption{Time-averaged concurrence plotted against $\kappa$ show a periodicity of $2\pi$. The initial state is an SCS with $\theta$ = 2.25 and $\phi$ = 2.0. The average was taken over 200 steps for the simulated plot and over 50 steps on IBM $vigo$.}
  \label{avg_conc}
\end{figure}

\begin{figure}[H]
  \centering
  \subfloat[]{\includegraphics[width=0.4\textwidth]{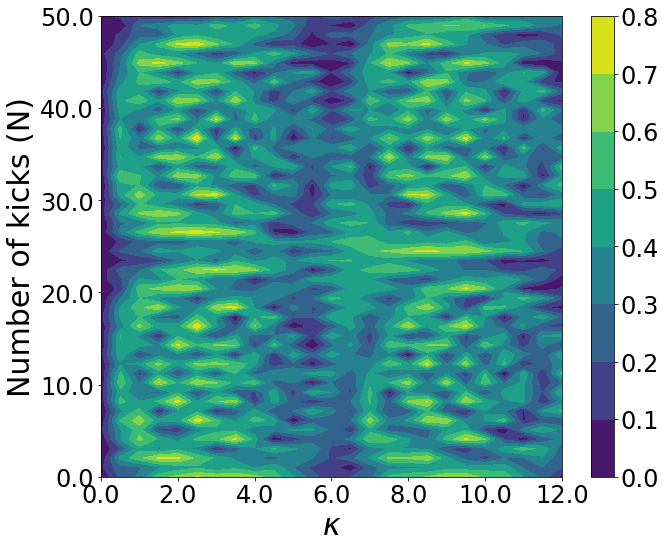}\label{contour_conc_exp}}
  \hfill
  \subfloat[]{\includegraphics[width=0.4\textwidth]{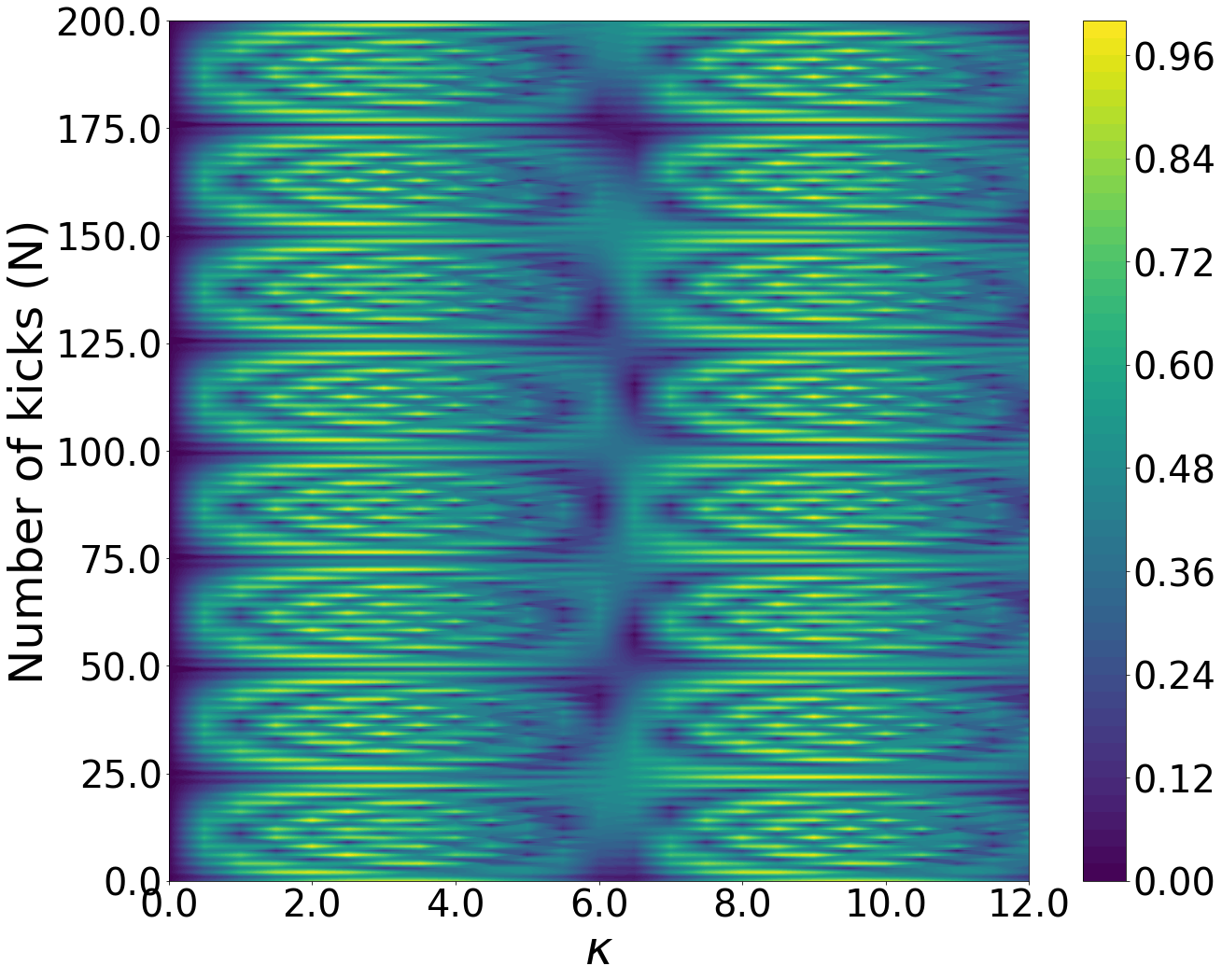}\label{contour_conc_simu}}
  \caption{ (a) Contour plot of concurrence over (a) 50 kicks for different values of $\kappa$ on IBMQ $vigo$ (b) 200 kicks for different values of kappa on IBMQ simulator. The initial state is an SCS with $\theta$ = 2.25 and $\phi$ = 2.0.}
\end{figure}

 The ability to explore a wide range of values of the chaoticity parameter $\kappa$ allows us to explore how chaos affects the dynamics. In particular, we study the effect of chaos (a classical phenomenon), on entanglement - a fundamentally quantum phenomenon. Here we consider the 2-qubit entanglement quantified by the concurrence as follows \cite{wootters}.  For a two-qubit density matrix $\rho$, the concurrence $C$ is defined to be
 
 \begin{equation}
    C=\max(0,\sqrt{\lambda_{1}}-\sqrt{\lambda_{2}}-\sqrt{\lambda_{3}}-\sqrt{\lambda_{4}}),
\end{equation}
where $\lambda_{i}$ are eigenvalues of  $\rho\Tilde{\rho}$ such that $\lambda_{4} \leq  \lambda_{3} \leq \lambda_{2} \leq \lambda_{1}$ and $0 \leq C \leq 1$, and $\Tilde{\rho}$ = $\sigma_{y} \otimes \sigma_{y} \rho^{*} \sigma_{y} \otimes \sigma_{y}$ (where $\sigma_{y}$ is Pauli matrix and $\rho^{*}$ is complex conjugate of $\rho$ in the standard basis). Concurrence is 0 for separable states and 1 for the maximally entangled Bell states.      

 Figure \ref{avg_conc} shows the time averaged 2-qubit concurrence as a function of $\kappa$ for an initial SCS centered at $\theta=2.25, \phi=2.0$. The experimental results from the IBMQ hardware show good agreement with the output from the IBM quantum circuit simulator, and both show the periodic behaviour of $\kappa$. The period of $2\pi$ agrees with the theoretical prediction of $2\pi j$ \cite{Santhanam}. To confirm the general periodicity of $\kappa$, we mapped the time evolution of the concurrence for 25 different values of $\kappa$ ranging from 0 to 12 to create a contour plot (Fig. 5). 
The periodic nature of the dynamics can be observed in the concurrence as we scan over either the number of kicks or $\kappa$, while holding the other variable constant. The periodicity of quantum dynamics in the 2-qubit QKT model was explored experimentally in one previous NMR study \cite{NMR}, which considered four different values of $\kappa$ and 8 kicks for each value of $\kappa$. Our work significantly expands the range of $\kappa$ and the number of kicks explored, and directly connects the periodicity to the observed entanglement dynamics.

\subsection{Entanglement and delocalization}

The periodicity in the concurrence, combined with the ability to implement a high number of kicks, can be exploited to generate detailed average concurrence plots. By averaging over multiple periods of concurrence in the number of kicks, we can reduce the error in the value of average concurrence. This allows more detailed observations of signatures of chaos in entanglements dynamics.

\begin{figure}[!h]
  \centering
  \subfloat[]{\includegraphics[width=0.28\textwidth]{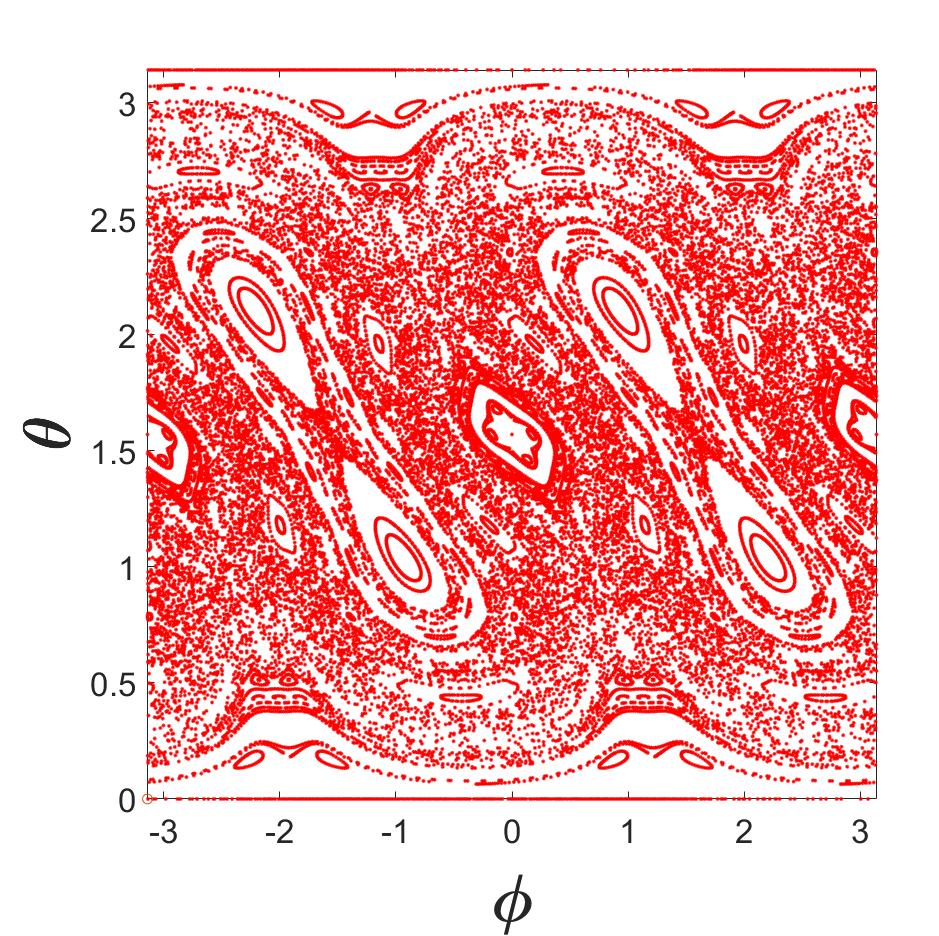}\label{classical_k2.5_1by1}}
  \hfill
  \subfloat[]{\includegraphics[width=0.33\textwidth]{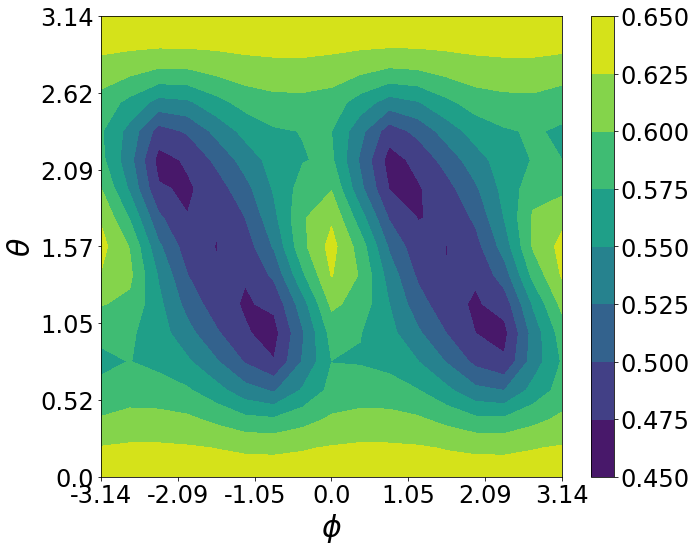}\label{avg_conc_contour_simu_2.5}}
  \hfill
  \subfloat[]{\includegraphics[width=0.33\textwidth]{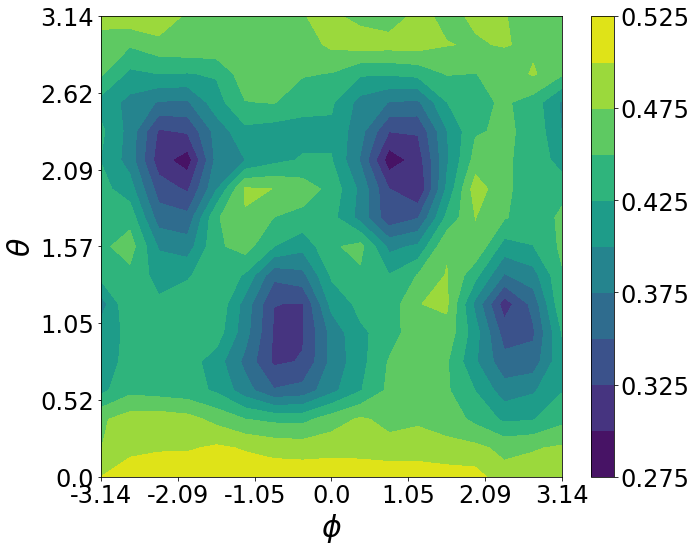}\label{avg_conc_contour_exp_2.5}}
  \caption{(a) Stroboscopic map, (b) average concurrence over 200 kicks on IBMQ simulator and (c) average concurrence over 50 kicks on IBMQ $vigo$ for 289 initial points and $\kappa$ = 2.5 .}
\end{figure}

A contour plot of time-averaged concurrence as function of $\theta$ and $\phi$ for $\kappa = 2.5$  
reflects the structures of the stroboscopic classical map as shown in Fig. 6.
Furthermore, our plots have enough resolution to observe that the chaotic regions of the classical phase space show intermediate concurrence values. The four prominently visible islands of low concurrence correspond to fixed points of the classical dynamics. These islands are clearly distinguishable on the hardware plot and the left-right symmetry is maintained as shown in Fig. \ref{avg_conc_contour_exp_2.5}.  Points $(J_x/j,J_y/j,J_z/j) = (1,0,0), (0,0,-1), (-1,0,0)$ and $(1,0,0)$, which constitute a period-4 orbit present in the classical dynamics of the system, show the highest values of average concurrence. 

We note the correspondence between this trend in average concurrence and the degree of delocalization of various initial states after evolution with the Floquet unitary. This degree of delocalization \cite{kumari_2019_quantumclassical} can be quantified by calculating the maximum overlap with respect to the set of minimum uncertainty spin coherent states. 

\begin{equation}\label{Oscs}
O_{\mathrm{SCS}}(|\psi(t)\rangle)=\max _{\mathrm{SCS}}|\langle\mathrm{SCS} \mid \psi(t)\rangle|. 
\end{equation} 

\begin{figure}[!h]
    \centering
    \includegraphics[scale=0.35]{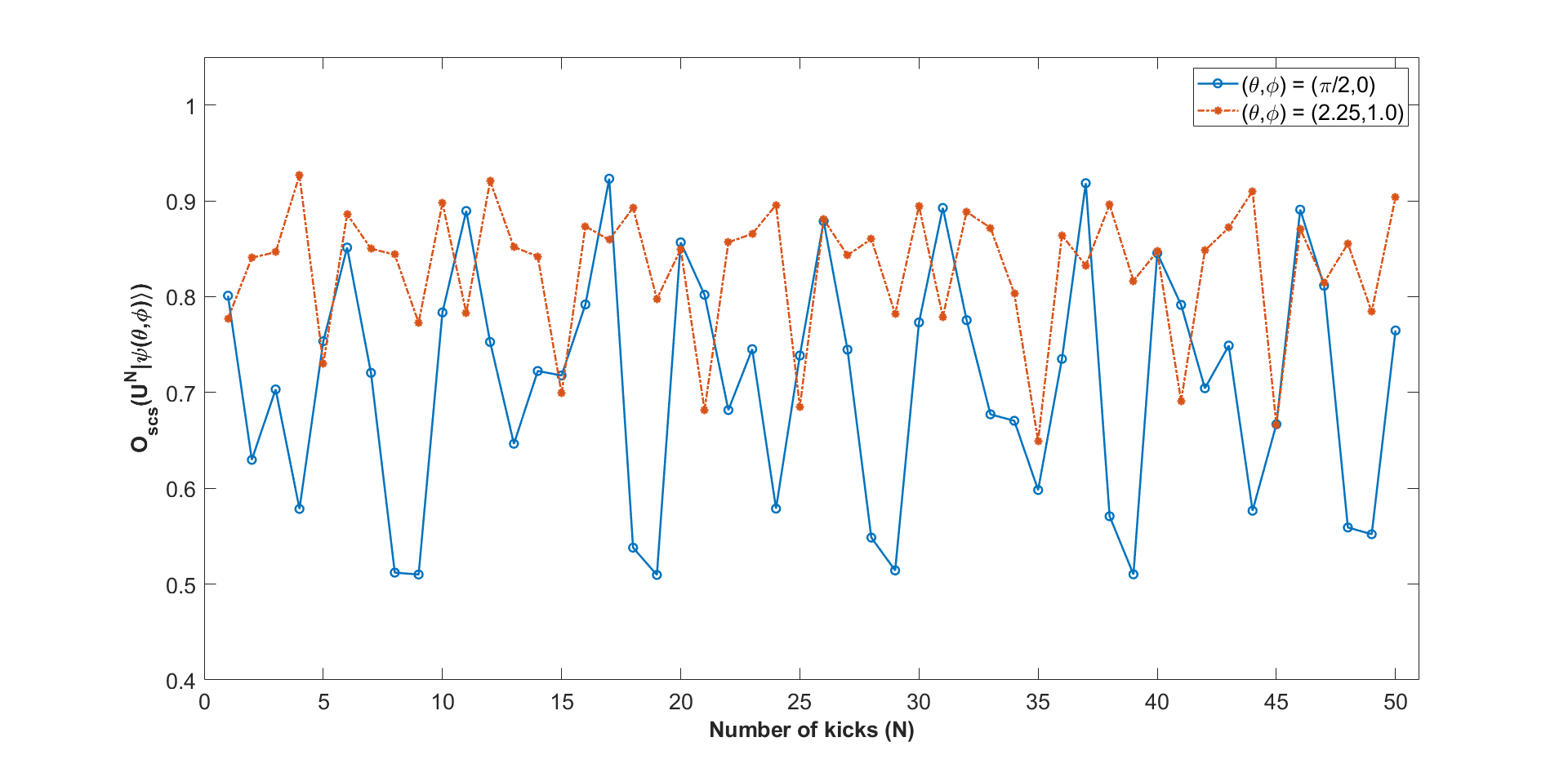}
    \caption{Evolution of $O_{\mathrm{SCS}}$ values for two different initial states for 50 kicks on on IBMQ $quito$. The evolution of initial state leading to higher concurrence ($(\theta,\phi)=(\pi/2,0)$) is more delocalized, i.e., has lower average $O_{\mathrm{SCS}}$ than that corresponding to lower concurrence ($(\theta,\phi)=(2.25,1)$).}
    \label{OSCS}
\end{figure}

Large values of $O_{\mathrm{SCS}}$ correspond to more localized states, as they indicate high overlap with spin coherent states. Delocalized states show low $O_{\mathrm{SCS}}$ values. The value of $O_{\mathrm{SCS}}$ for two different states, one in the high concurrence region and one in the low concurrence region are plotted against number of steps in Fig. \ref{OSCS}, and show the connection between entanglement and delocalization. This is the first observation of this connection in the deepest quantum regime and confirms previous theoretical predictions~\cite{NMR},\cite{pattanayak},\cite{dogra_google}.


\section{Summary and discussion} \label{sec 5}


In this work, we have proposed a quantum circuit-based approach to simulate and explore quantum chaos, and demonstrated its advantages over existing methods. The proposed method can be applied in general to any periodically driven finite dimensional quantum system. In our study, IBM's 5-qubit open access quantum chip ($vigo$) was used as the experimental platform to implement the proposed approach for the 2-qubit quantum kicked top (QKT).  The Hamiltonian of the QKT can be exactly expressed in terms of qubits since it is a finite-dimensional quantum system. Therefore, its evolution operator can be  decomposed into quantum gates. Traditionally, experimental studies of quantum chaos have applied the same set of operations $n$ times to explore time evolution.
Here, we decomposed the unitary evolution operator for $n$ kicks, $U^n$, into elementary quantum gates. This results in a fixed number of operations implementing the QKT evolution for any number of kicks. This hybrid combination of classical processing and quantum computing opens up the ability to perform high fidelity experimental studies of quantum chaos in new parameter regimes. 

Since the value of the chaoticity parameter $\kappa$ only determines the parameters of unitary rotations in the quantum circuit, and since the single qubit rotation errors are independent of the parameters, we were able to experimentally study chaotic dynamics over a wider range of $\kappa$ and kick number compared to previous studies. By taking advantage of the high fidelity obtained for both a large number of kicks and arbitrary $\kappa$ values, we experimentally demonstrated the periodicity of entanglement with time and $\kappa$ with high accuracy. 
Our studies also clearly showed signature of chaos in the contour plot of average 2-qubit concurrence despite being in the deeply quantum regime. Furthermore, we reported the first observation of the correspondence between average entanglement and delocalization in the 2-qubit QKT. 

Our results demonstrate the advantages of circuit-based NISQ devices for exploring fundamental questions in quantum information and quantum chaos despite their noise and scale limitations.
Previous studies 
\cite{PhysRevE.62.3504},\cite{PhysRevE.62.6366} have noted that chaos could influence the efficient and stable operation of quantum computers.
In \cite{Berke2020TransmonPF}, it was shown that chaos affects the balance between the disorder that maintains the stability of qubits and nonlinear resonator couplings that is used to manipulate interactions. This plays an integral role in future transmon device engineering. The gate-based circuit model of QKT can be used as an efficient tool for studying these effects in different situations.   


\section*{Acknowledgements}
We thank M. Kumari, R. Mann, A. Mashatan and N. Lutkenhaus for useful discussions.
This work was supported by the  Natural Sciences and Engineering Research Council of Canada. Wilfrid Laurier University is located in the traditional territory of the Neutral, Anishnawbe and Haudenosaunee peoples. We thank them for allowing us to conduct this research on their land.

\printbibliography
\end{document}